\documentclass[doublecol]{epl2}
\usepackage{cite}
\usepackage{epsf}
\usepackage{epsfig}
\usepackage{epstopdf}
\usepackage{float}
\usepackage{amssymb}
\usepackage{amsfonts}
\usepackage{color}
\usepackage{url}

\def\k{{{\bf k}}}

\def\g{{\bf{g}}}

\def\beq{\begin{equation}}
\def\eeq{\end{equation}}
\def\beqa{\begin{eqnarray}}
\def\eeqa{\end{eqnarray}}

\def\g0{{\gamma_0}}

\def\prb{Phys.\ Rev.\ B, }
\def\prl{Phys.\ Rev.\ Lett.,\ }
\def\jpcm{J.\ Phys.\ Condens.\ Matter, }

\def\RMP{Rev.\ Mod.\ Phys.,\ }
\def\JPSJ{J.\ Phys.\ Soc.\ Japan,}

\newcommand\bear{\begin{eqnarray}}
\newcommand\eear{\end{eqnarray}}
\newcommand\bea{\begin{align}}
\newcommand\ena{\end{align}}
\newcommand{\email}[1]{\texttt{#1}}
\usepackage{comment}
\usepackage{setspace}

\title{Proximity effects in a superconductor-normal metal bilayer system}
\author{Naushad Ahmad Kamar and N.\ S.\ Vidhyadhiraja}
\institute{Theoretical Sciences Unit\\Jawaharlal Nehru Centre for Advanced Scientific Research,\\
 Jakkur, Bangalore 560 064, India}
 \abstract{In this letter, we have studied the effects of proximity of a superconductor to a normal
metal. The system, represented by a  bilayer attractive Hubbard model,
is investigated using layer-dynamical mean field theory  and iterated perturbation theory for superconductivity as an impurity solver. The bilayer system comprises a superconducting and a normal metallic layer, connected by 
an inter-planar hopping ($t_\perp$). It is found that superconductivity is induced in the normal layer for small $t_\perp$. With increasing inter-layer hopping, the bilayer system undergoes two transitions: a first order transition to a normal metallic phase, and subsequently a continuous
crossover to a band insulator.}
\begin{document}
\maketitle
\section{Introduction}

The physical proximity of two or more distinct phases of matter can generate exotic
phenomena at the interface. Spectacular examples of such phenomena include formation
of a depletion layer at p-n junctions~\cite{4Gerold}, giant  magnetoresistance effect in alternating
ferromagnetic and non-magnetic metallic heterostructures~\cite{4Fere}, formation of a two-dimensional
electron gas at the interface of a band insulator and a Mott insulator~\cite{4William}. 
Similarly,  when a superconductor is brought into contact with a normal metal or a ferromagnet,  then many interesting phenomena take place at the interfaces such as  Andreev reflection~\cite{4Andreev}, induction of superconducting (SC) correlations in
 normal metal, triplet pairing at the interface of superconductor and ferromagnet etc. 
These phenomena are collectively known as proximity effects~\cite{4gennes}.

There are many experimental realizations of the superconductor-normal metal (SN) interface, such 
as Nb/Au~\cite{4Kim,4Hiroki} and $\mathrm{NbSe_2}$/Au~\cite{4Truscott}. Similarly, proximity effects in a superconductor-ferromagnet(SF) interface
have been realized in Nb/Fe~\cite{4Kawaguchi,4Verbanck,4Muhge}, Nb/Gd~\cite{4Strunk,4Jiang}, V/Fe~\cite{4Wong,4Koorevaar,4Tagirov}, V/$\mathrm V_{1-x}\mathrm {Fe}_x$~\cite{4Aarts}, Pb/Fe~\cite
{4Lazar}, Co/Al ~\cite{4Goto}, and 
 $\mathrm {YBa_2Cu_3O_{7-\delta}}$/ $ \mathrm{La_{0.7}Co_{0.3}MnO_3}$~\cite{4Sefrioui} etc. At an SN or an SF interface, it is observed that the superconductivity is induced
 in the metallic/ferromagnetic (M/FM) layer because of leakage of Cooper pairs from superconductor to M/FM~\cite{4Wong}. The superconducting transition temperature decreases with increasing the thickness of the non-SC second layer.
The magnetic moments in ferromagnet break Cooper pairs inside the ferromagnet because of breaking of time reversal symmetry and hence the superconducting critical temperature decreases as function of thickness of ferromagnetic layer faster than in an SN layer~\cite{4Wong}. 

The proximity of a superconductor to a normal metal is, theoretically,
a well studied problem~\cite{4Werthamer, 4gennes, 4Fominov}, albeit 
 with static mean field theories. 
The static mean field theories, however, do not incorporate dynamical fluctuations,
which are very important in accessing the true ground state. Nevertheless,
theoretical studies of the SN interface that go beyond static mean field theory
are scant.
In a recent theoretical study, an SN interface has been represented by a 
bilayer attractive Hubbard model on a square lattice of finite size. The
model has been solved using Quantum Monte Carlo (QMC) as well as
 Bogoliubov-de Gennes mean field (BdGMF) approximation~\cite{4Zujev}.

In this letter, we have studied the bilayer attractive Hubbard model (AHM) 
 by combining dynamical mean field theory (DMFT)~\cite{4Antoine,4Kotliar, 4Dieter} and  iterated perturbation theory for superconductivity (IPTSC)~\cite{4Arti,4Kawa, 4Naushad} as an impurity solver. The IPTSC method is perturbative, but is known to benchmark excellently when compared to 
 results from numerical renormalization group for the bulk AHM~\cite{4Bauer}. Although the target system is the same as that
 of Ref.~\cite{4Zujev}, our method allows us to study the thermodynamic limit, thus avoiding any finite-size effects. Moreover, since the IPTSC has been implemented on a real frequency axis, we have been able to investigate 
 spectral dynamics in detail without facing the ill-posed problem of analytic
 continuation.  The various phases of the system 
 as a function of increasing $t_\perp$ have been explored. 
The paper is structured as follows: In the following section, we outline the model and the formalism used. Next, we present our results for the spectra,
order parameter and phase transitions. We conclude in the final section.
 
\section{Model and Method}
 We consider a single band bilayer attractive Hubbard model (AHM), which 
may be represented by the following Hamiltonian:
\begin{eqnarray}
\hspace*{-0cm}{\cal H}=
\sum_{i\sigma,l=1}^{l=2}\epsilon_{l} c^{\dagger}_{il\sigma}c^{\phantom{\dag}}_{il\sigma}-\sum_{<ij\sigma>,l=1}^{l=2}t_l[c^{\dagger}_{il\sigma}c^{\phantom{\dag}}_{jl\sigma}+h.c]- \nonumber \\
\sum_{i,l=1}^{l=2}|U_l|(n_{il\uparrow}-\frac{1}{2})(n_{il\downarrow }-\frac{1}{2})- \nonumber \\
\sum_{i\sigma,l=1}^{l=2}\mu_lc^{\dagger}_{il\sigma}c^{\phantom{\dag}}_{il\sigma}-t_\perp\sum_{i\sigma}[c^{\dagger}_{i1\sigma}c^{\phantom{\dag}}_{i2\sigma}+h.c]
\label{eq:4eq1}
\end{eqnarray}
where ${c}_{il\sigma}$ annihilates an electron with spin $\sigma$ on the 
$i^{\rm th}$ lattice site in the $l{\rm th}$  plane. The local occupancy
is determined by the operator, ${n}_{il\sigma}=c^{\dagger}_{il\sigma}c^{\phantom{\dagger}}_{il\sigma}$.
The indices $ i, j$ run over the lattice sites in each plane and $l$ is 
a plane index. $t_\perp$ is inter-planar hopping and $t_l$ is intra-planar hopping in the $l_{th}$ plane; $\epsilon_l$ is the site energy of 
$l_{th}$ plane.

To take superconductivity into account, we use a four component Nambu spinor, which is defined as
\begin{equation}
\hspace*{-1cm} \Psi^\dag_{k} =
 \left[ \begin{array}{llll}
c_{k1\uparrow}&
c_{-k1\downarrow}^{\dagger} &
c_{k2\uparrow} &
c_{-k2\downarrow}^{\dagger}
\end{array} \right]
\label{eq:4eq2}
\end{equation}
The matrix Green's function is given by
\begin{equation}
\hat G(\vec k,\tau)=-<T_{\tau}\Psi^{\phantom{\dag}}(\vec k,\tau)\Psi^{\dagger}(\vec k,0)>
\label{eq:4eq3}
\end{equation}
where 1 and 2 label the planes and $\vec k$ is momentum quantum number.
The Green's function in absence of interaction ($U_1=U_2=0$) is given by
\begin{equation}
\hspace*{0.cm} \hat{G}_0(\vec k,\omega) =
 \left[ \begin{array}{llll}
{ \omega^{+}-\bar{\epsilon}_1(\vec k)} & { 0} \\ &  { t_{\perp}} & { 0}\\
 {0 }&{ \omega^{+}+\bar{\epsilon}_1(\vec k)} \\ &{  0}  &{-t_{\perp}}\\
 {  t_{\perp}}& { 0} \\ &{\omega^{+}-\bar{\epsilon}_2(\vec k)}&  {  0}\\
 { 0} &{ -t_{\perp}} \\ &{  0}&{\omega^{+}+{\bar{\epsilon}}_2(\vec k)}
\end{array} \right]^{-1}
\label{eq:4eq4}
\end{equation}
where $\bar{\epsilon}_l(\vec k)=\epsilon_l(\vec k)-\mu_l+\epsilon_l$
and
$\epsilon_l(\vec k)$  is the 
dispersion relation for the $l^{\rm th}$ plane. Then, the interacting Green's function is obtained by using the Dyson's equation
\begin{equation}
\hat{G}^{-1}(\vec k,\omega) =\hat{G}^{-1}_0(\vec k,\omega)-\hat\Sigma(\omega)
\label{eq:4eq5}
\end{equation}
where $\hat\Sigma(\omega)$ is self-energy matrix, and is given by
\begin{equation}
\hspace*{0cm} \hat{\Sigma}(\omega) =
 \left[ \begin{array}{llll}
  \Sigma_1(\omega) &S_1(\omega)&0&0\\
 S_1(\omega) &-\Sigma^{\ast}_1(-\omega)&0&0\\
 0&0&\Sigma_2(\omega)&S_2(\omega)\\
 0&0& S_2(\omega)&-\Sigma^{\ast}_2(-\omega)
\end{array} \right]
\label{eq:4eq6}
\end{equation}
where $\Sigma_1(\omega)$, $ S_1(\omega)$, $\Sigma_2(\omega)$ and $S_2(\omega)$ are the normal and anomalous self-energies of planes 1 and 2  respectively.
In order to calculate the local self-energies, 
we use the IPTSC\cite{4Arti} as
 an impurity solver. In the IPTSC method, based on second order perturbation theory,
the self-energies are given by the following ansatz:
\begin{eqnarray}
\Sigma_{1}(\omega)&=&-U_1\frac{n_1}{2}+A_1\Sigma^{(2)}_{1}(\omega)
\label{eq:4eq7} \\
S_{1}(\omega)&=&-U_1\Phi_1 + A_1S^{(2)}_{1}(\omega)
\label{eq:4eq8}\\
\Sigma_{2}(\omega)&=&-U_2\frac{n_2}{2}+A_2\Sigma^{(2)}_{2}(\omega)
\label{eq:4eq9} \\
S_{2}(\omega)&=&-U_2\Phi_2 + A_2S^{(2)}_{2}(\omega)\,
\label{eq:4eq10}
\end{eqnarray}
where the local filling $n_1$, $n_2$ and order parameter $\Phi_1$, $\Phi_2$  are given by
\begin{eqnarray}
n_1&=&-\frac{2}{\pi}\int_{-\infty}^{\infty}d\omega\,{\rm Im}(G_{11}(\omega))\,f(\omega)
\label{eq:4eq11}\\
\Phi_1&=&\int_{-\infty}^{\infty}d\omega\frac{-{\rm Im}(G_{12}(\omega))}{\pi}
f(\omega)
\label{eq:4eq12}\\
n_2&=&-\frac{2}{\pi}\int_{-\infty}^{\infty}d\omega\,{\rm Im}(G_{33}(\omega))\,f(\omega)
\label{eq:4eq13}\\
\Phi_2&=&\int_{-\infty}^{\infty}d\omega\frac{-{\rm Im}(G_{34}(\omega))}{\pi}
f(\omega)
\label{eq:4eq14}
\end{eqnarray}
and $f(\omega)=\theta(-\omega)$ is the Fermi-Dirac distribution function at zero temperature.
In the ansatz above, (equations~(\ref{eq:4eq4} and \ref{eq:4eq5})), the second order self-energies are given by
\begin{eqnarray}
\Sigma^{(2)}_{1}(\omega)&=&U_1^2\int_{-\infty}^{\infty}\prod_{j=1}^{3}d\omega_{j}\frac{g_{11}(\omega_1,\omega_2,\omega_3)N(\omega_1,\omega_2,\omega_3)}{\omega^{+}-\omega_1+\omega_2-\omega_3} \nonumber \\
 \nonumber \\
S^{(2)}_{1}(\omega)&=&U_1^2\int_{-\infty}^{\infty}\prod_{j=1}^{3}d\omega_{j}\frac{g_{21}(\omega_1,\omega_2,\omega_3)N(\omega_1,\omega_2,\omega_3)}{\omega^{+}-\omega_1+\omega_2-\omega_3}  \nonumber \\
 \nonumber
\Sigma^{(2)}_{2}(\omega)&=&U_2^2\int_{-\infty}^{\infty}\prod_{j=1}^{3}d\omega_{j}\frac{g_{12}(\omega_1,\omega_2,\omega_3)N(\omega_1,\omega_2,\omega_3)}{\omega^{+}-\omega_1+\omega_2-\omega_3} \nonumber \\
{\mbox{and}} \nonumber \\
S^{(2)}_{2}(\omega)&=&U_2^2\int_{-\infty}^{\infty}\prod_{j=1}^{3}d\omega_{j}\frac{g_{22}(\omega_1,\omega_2,\omega_3)N(\omega_1,\omega_2,\omega_3)}{\omega^{+}-\omega_1+\omega_2-\omega_3} \nonumber \\
\label{eq:4eq15}
\end{eqnarray}
where 
\begin{eqnarray}
 N(\omega_1,\omega_2,\omega_3)&=&f(\omega_1)f(-\omega_2)f(\omega_3)+       \nonumber \\ f(-\omega_1)f(\omega_2)
f(-\omega_3) \nonumber \\
g_{11}(\omega_1,\omega_2,\omega_3)&=&\tilde{\rho}_{11}(\omega_1)\tilde{\rho}_{22}(\omega_2)\tilde{\rho}_{22}(\omega_3)- \nonumber \\ \tilde{\rho}_{12}(\omega_1)\tilde{\rho}_{22}(\omega_2)\tilde{\rho}_{12}(\omega_3)\nonumber \\
g_{21}(\omega_1,\omega_2,\omega_3)&=&\tilde{\rho}_{12}(\omega_1)\tilde{\rho}_{12}(\omega_2)\tilde{\rho}_{12}(\omega_3)- \nonumber \\ \tilde{\rho}_{11}(\omega_1)\tilde{\rho}_{12}(\omega_2)\tilde{\rho}_{22}(\omega_3)\nonumber \\
g_{12}(\omega_1,\omega_2,\omega_3)&=&\tilde{\rho}_{33}(\omega_1)\tilde{\rho}_{44}(\omega_2)\tilde{\rho}_{44}(\omega_3)- \nonumber \\ \tilde{\rho}_{34}(\omega_1)\tilde{\rho}_{44}(\omega_2)\tilde{\rho}_{34}(\omega_3)\nonumber \\
g_{22}(\omega_1,\omega_2,\omega_3)&=&\tilde{\rho}_{34}(\omega_1)\tilde{\rho}_{34}(\omega_2)\tilde{\rho}_{34}(\omega_3)- \nonumber \\ \tilde{\rho}_{33}(\omega_1)\tilde{\rho}_{34}(\omega_2)\tilde{\rho}_{44}(\omega_3)
\label{eq:4eq16}
\end{eqnarray}
and the spectral functions $\tilde{\rho}^i_{\alpha\beta}$, $\alpha,\beta=1,4$ are given by the imaginary part
of the `Hartree-corrected' host Green's function, namely
${\hat{\tilde{\rho}}}(\omega)=-{\rm Im}\hat{\cal {G}}(\omega)/\pi$. The latter is given by  
\begin{equation} 
\hspace*{-0.0cm} \hat{\cal{G}}(\vec k, \omega) =
\left[ \begin{array}{llll}
{ \scriptstyle\omega^{+}-{\bar{\epsilon}}_1(\vec k)+U_1\frac{n_1}{2}} & {\scriptstyle U_1\Phi_1} \\ & {\scriptstyle t_{\perp}} & {\scriptstyle 0}\\
 {\scriptstyle U_1\Phi_1 }&{\scriptstyle \omega^{+}+{\bar{\epsilon}}_1(\vec k)-U_1\frac{n_1}{2}} \\ &{ \scriptstyle 0}&{\scriptstyle-t_{\perp}}\\
 { \scriptstyle t_{\perp}}& {\scriptstyle 0} \\ &{ \scriptstyle \omega^{+}-{\bar{\epsilon}}_2(\vec k)+U_2\frac{n_2}{2}}&{\scriptstyle U_2\Phi_2}\\
 { \scriptstyle 0} &{ \scriptstyle-t_{\perp}}  &{ \scriptstyle U_2\Phi_2}\\&{\scriptstyle\omega^{+}+{\bar{\epsilon}}_2(\vec k)-U_2\frac{n_2}{2}}
\end{array} \right]^{-1}
\label{eq:4eq17}
\end{equation}
\begin{eqnarray}
\hat{\cal {G}}(\omega)=\sum_{\vec k}\hat{\cal {G}}(\vec k,\omega)
\nonumber \\
\hat{ {G}}(\omega)=\sum_{\vec k}\hat{ {G}}(\vec k,\omega)
\label{eq:4eq18}
\end{eqnarray}
Finally the coefficient $A_l$, which
is determined by the high frequency limit, in the IPTSC ansatz equations~(\ref{eq:4eq7},
  \ref{eq:4eq8}, \ref{eq:4eq9}, and \ref{eq:4eq10}),  is given by
\begin{equation}
A_l=\frac{\frac{n_l}{2}(1-\frac{n_l}{2})-\Phi_l^2}{{\frac{n_{0l}}{2}(1-\frac{n_{0l}}{2})-\Phi_{0l}^2}}
\label{eq:4eq19}
\end{equation}
where the pseudo order-parameter $\Phi_{0l}$ and the pseudo occupancy $n_{0l}$,
are given by 
\begin{eqnarray}
& n_{01}=&2\int_{-\infty}^{\infty}d\omega \tilde{\rho}_{11}(\omega) f(\omega) \nonumber \\
&\Phi_{01}=&\int_{-\infty}^{\infty}d\omega \tilde{\rho}_{12}(\omega) f(\omega)\nonumber \\
& n_{02}=&2\int_{-\infty}^{\infty}d\omega \tilde{\rho}_{33}(\omega) f(\omega) \nonumber \\
{\mbox {and}}\;\;&\Phi_{02}=&\int_{-\infty}^{\infty}d\omega \tilde{\rho}_{34}(\omega) f(\omega)\,.
\label{eq:4eq20}
\end{eqnarray} 

\subsection{ Numerical Algorithm}
The algorithm to solve above equations is given below:
\begin{enumerate}
\item Guess $\hat \Sigma$ and calculate  $\hat G$ by using equations~\ref{eq:4eq4} and \ref{eq:4eq5}.
\item Then, by using equations~\ref{eq:4eq11},\ref{eq:4eq12}, \ref{eq:4eq13}, \ref{eq:4eq14} and \ref{eq:4eq17}, calculate effective medium propagator $\hat{\cal G}(\omega)$.
\item By using effective medium propagator, calculate the new self-energy matrix.
\item If initial and final self-energy matrices have converged within a desired accuracy, then stop, else feedback this new self-energy to step 1.
\end{enumerate}
The results obtained using the above-mentioned procedure will be denoted as
IPTSC. We have also carried out mean-field calculations by ‘turning off’ the dynamical
self-energies in equations~\ref{eq:4eq7}, \ref{eq:4eq8}, \ref{eq:4eq9} and \ref{eq:4eq10}. These results will be denoted as BdGMF.

\section{Results and discussion}

In this letter, we have considered plane-1 to be interacting $(U_1\neq0)$ and plane-2 to be non-interacting $(U_2=0)$. Both planes are at half filling
which is fixed by taking $\mu_1= \mu_2=\epsilon_1=\epsilon_2=0.0 $. 
We have taken $t_1= t_2=1$ as an energy unit, and $U_1=U=2.0$. 
Both layers are Bethe lattices of infinite connectivity, hence the $\k$-summations in equations 18 may be converted to a density of states integral
as done in Ref~\cite{4Naushad}.

\subsection{Spectral functions}
To understand the proximity effect, we have analyzed the spectral functions of both interacting ($U\neq0$) and non-interacting ($U=0$) planes for different values of $t_\perp$. 
\begin{figure}[ht]
\centering
\includegraphics[scale=0.425,clip]{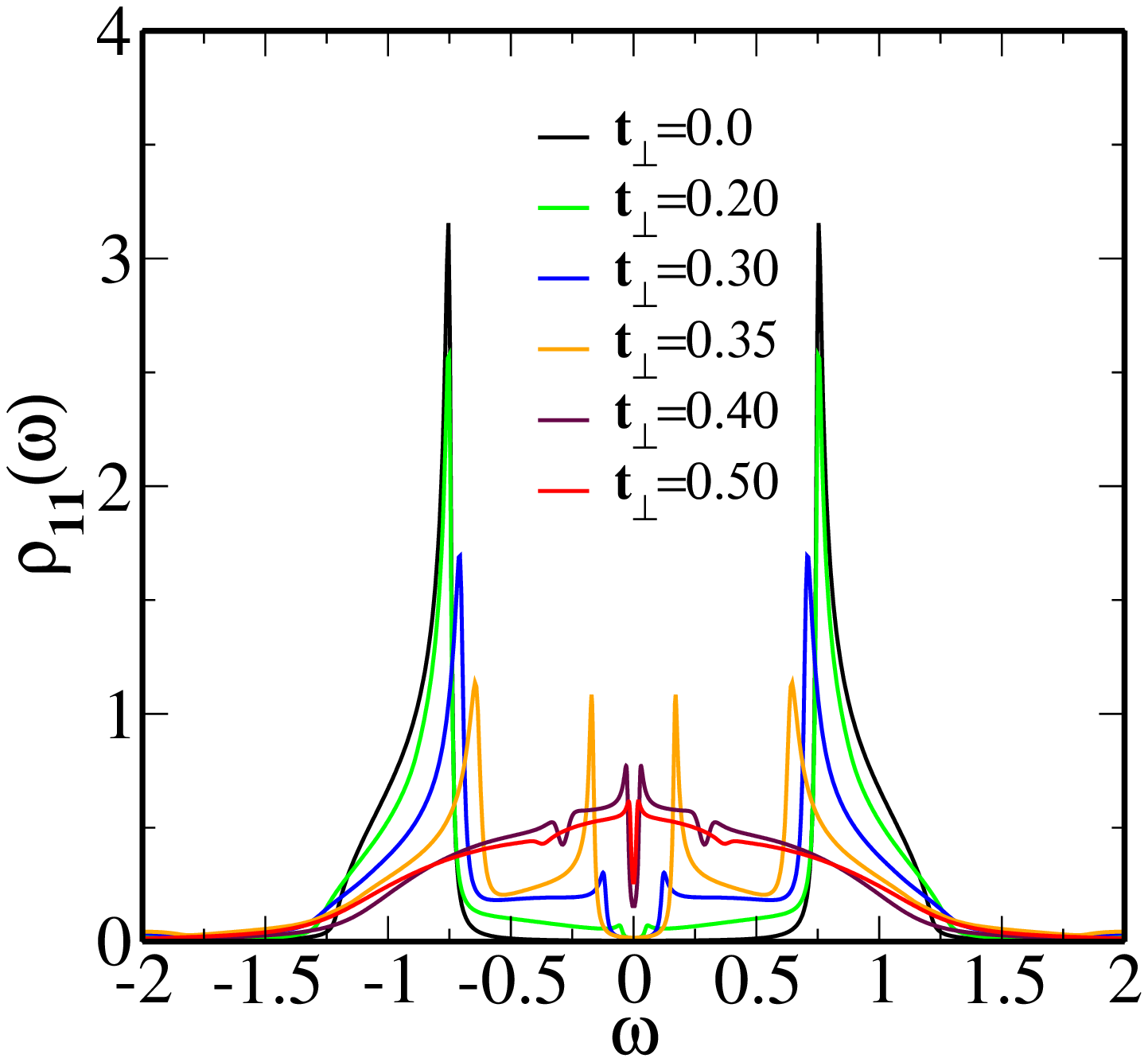}
\includegraphics[scale=0.425,clip]{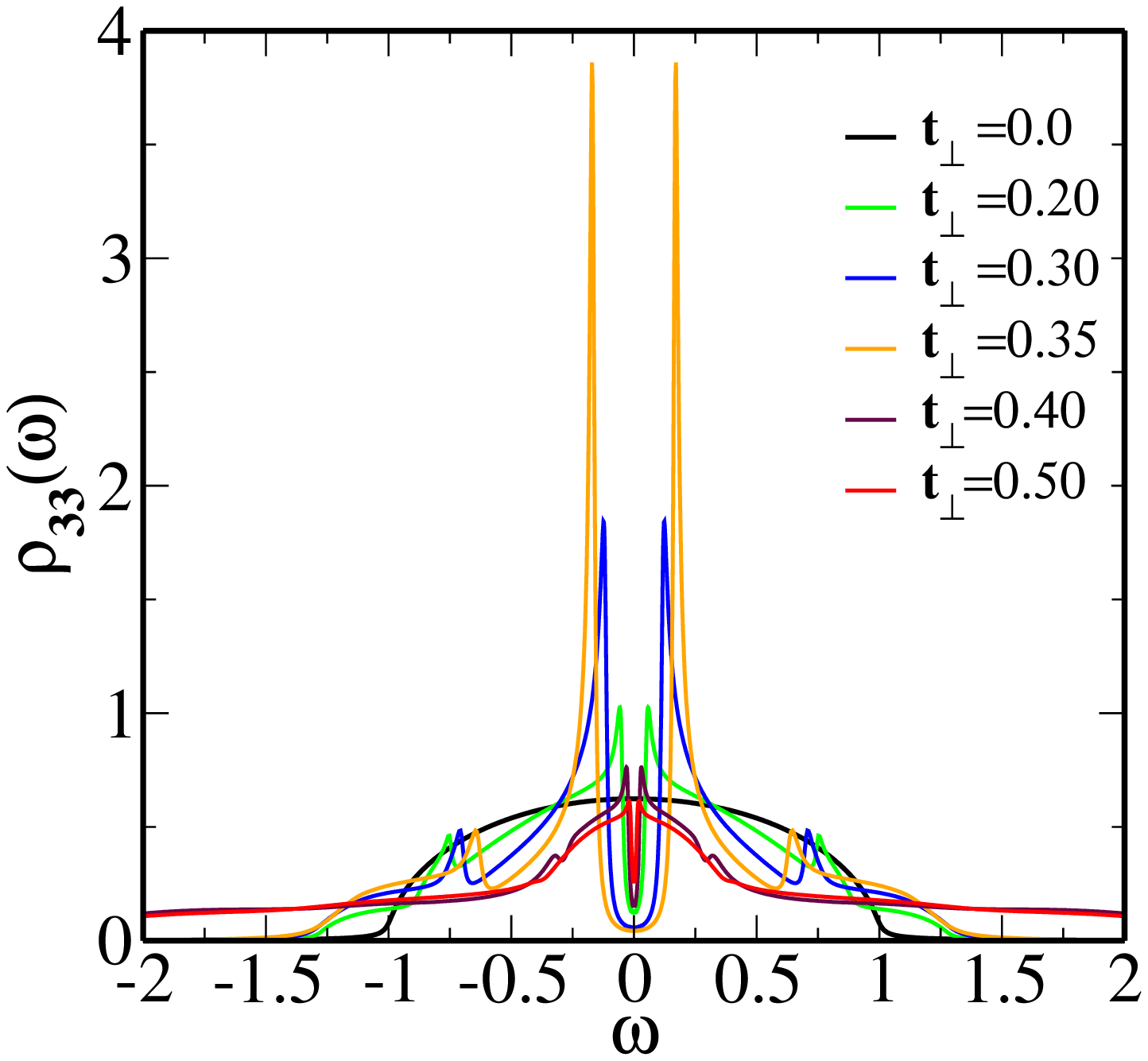}
\caption{Upper panel: interacting layer spectral function ; Lower panel: non-interacting layer spectral function at U=2.0, and at $\langle n_1\rangle=\langle n_2\rangle=1.0$ for different values of $t_\perp$.}
\label{fig:4fig1}
\end{figure}

In figure~\ref{fig:4fig1}, the diagonal component of the spectral function as function of $\omega$ is shown for different values of  $t_\perp$. The upper panel represents the spectral function of interacting
 plane while lower panel represents the spectral function of the non-interacting plane. In upper panel, there is a sharp coherence peak at the gap edge of
  spectral function at $t_\perp=0.0$, which is a characteristic of s-wave superconductivity. The spectral weight of the coherence peak decreases with increasing $t_\perp$, indicating that superconducting order is decreasing
in interacting plane because of proximity to non-interacting plane. In the 
lower panel, at $t_\perp=0.0$, spectral function is semi-elliptic because it
is basically the non-interacting spectral function of an infinite-dimensional Bethe lattice. With increasing $t_\perp$, the non-interacting spectral function becomes gapped. As will be discussed later, this gap is due to induction of
superconductivity in the non-interacting layer caused by the proximity
to the SC layer. At $t_\perp=0.20$, there is a sharp coherence peak at the gap edge and weight in the peak increases with increasing $t_\perp$ reaching a 
maximum at $t_\perp=0.35$.
\begin{figure}[t]
\centering
\includegraphics[scale=0.425,clip]{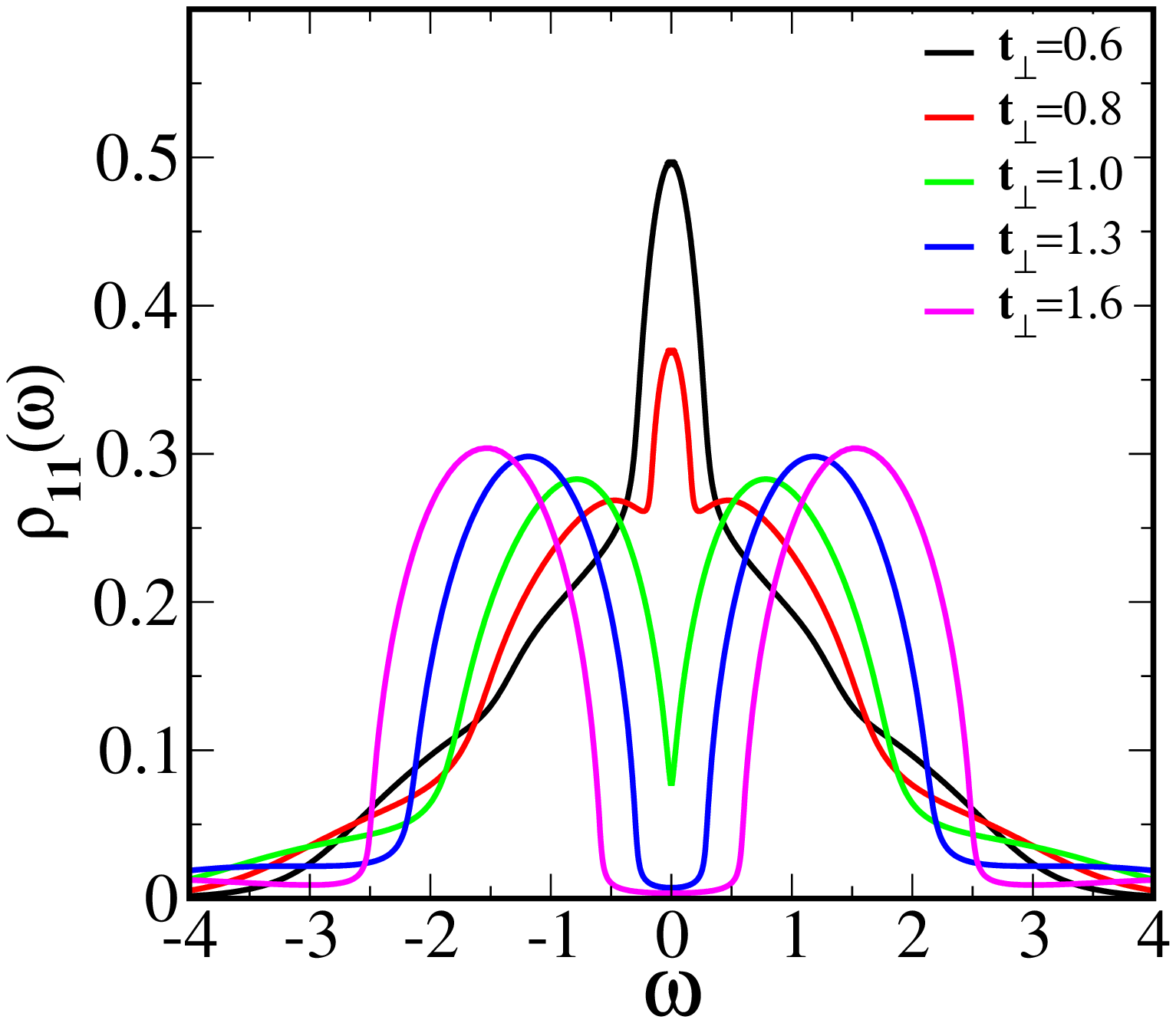}
\includegraphics[scale=0.425,clip]{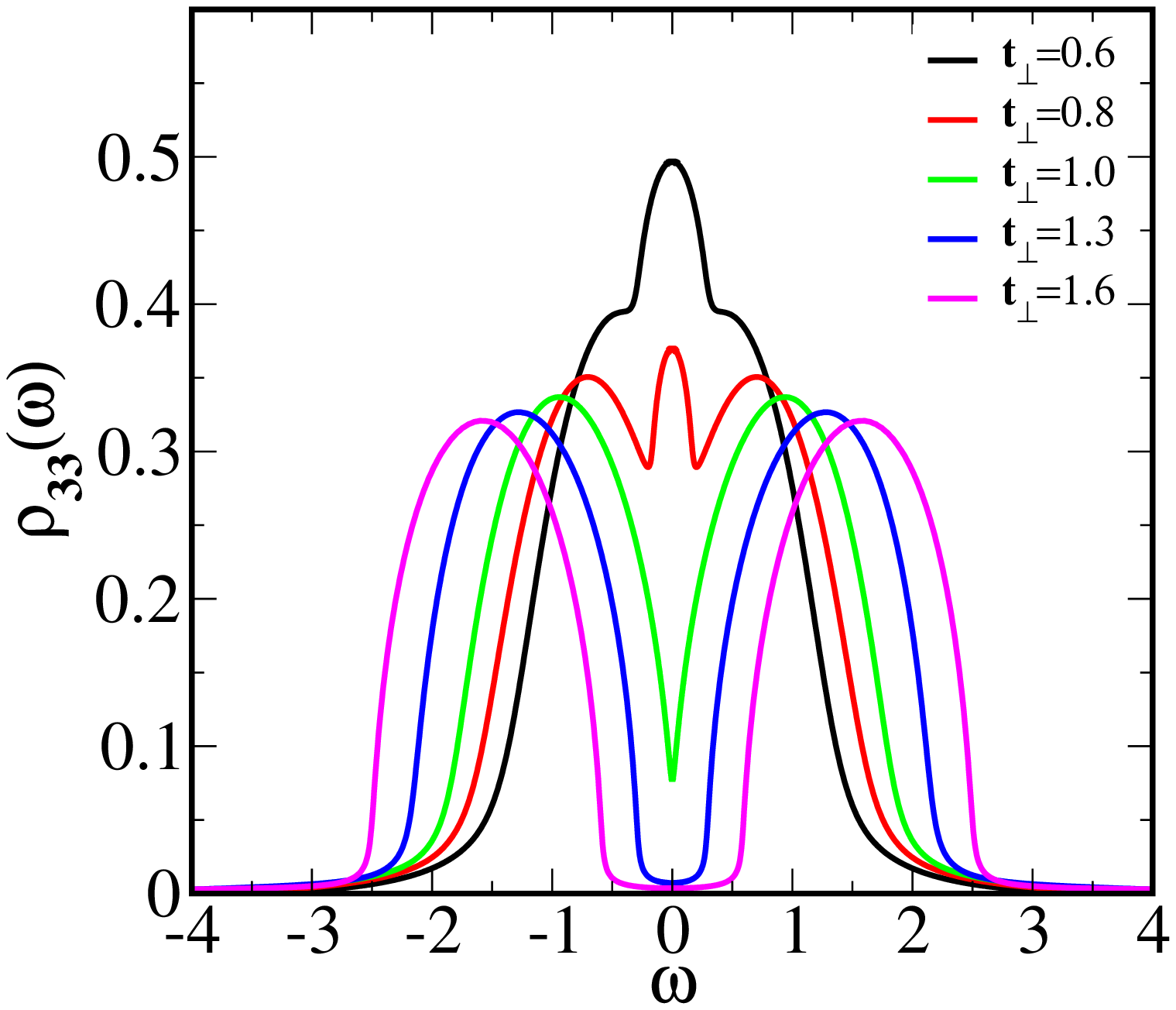}
\caption{Upper panel: interacting layer spectral function ; Lower panel: non-interacting layer spectral function at U=2.0, and at $\langle n_1\rangle=\langle n_2\rangle=1.0$ for different values of $t_\perp$.}
\label{fig:4fig2}
\end{figure}
In figure~\ref{fig:4fig2}, the diagonal component of the spectral functions of 
both the interacting and non-interacting planes are shown for higher values of $t_\perp$. The upper panel represents the spectral function of the interacting plane and lower panel represents the spectral function of the non-interacting plane. At $t_\perp=0.60$, the coherence peak at gap edge in both interacting and interacting planes completely vanishes and both planes become metallic. Further increasing $t_\perp$, both interacting and non-interacting spectral functions become gapped. The nature of this gap, that occurs at large $t_\perp$
 will be discussed later. The spectral gap increases with increasing $t_\perp$.

\begin{figure}[]
\centering
\includegraphics[scale=0.425,clip]{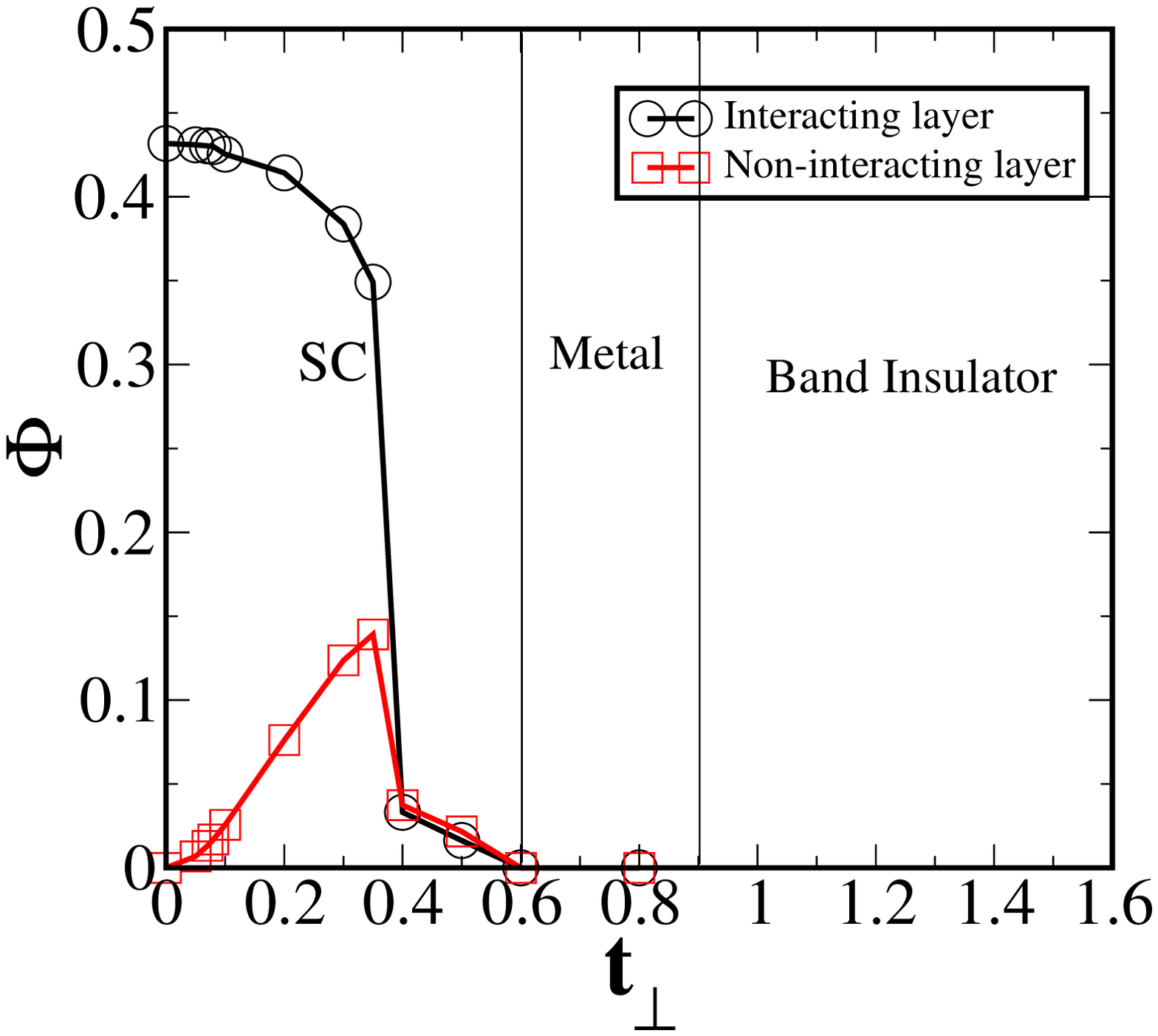}
\includegraphics[scale=0.425,clip]{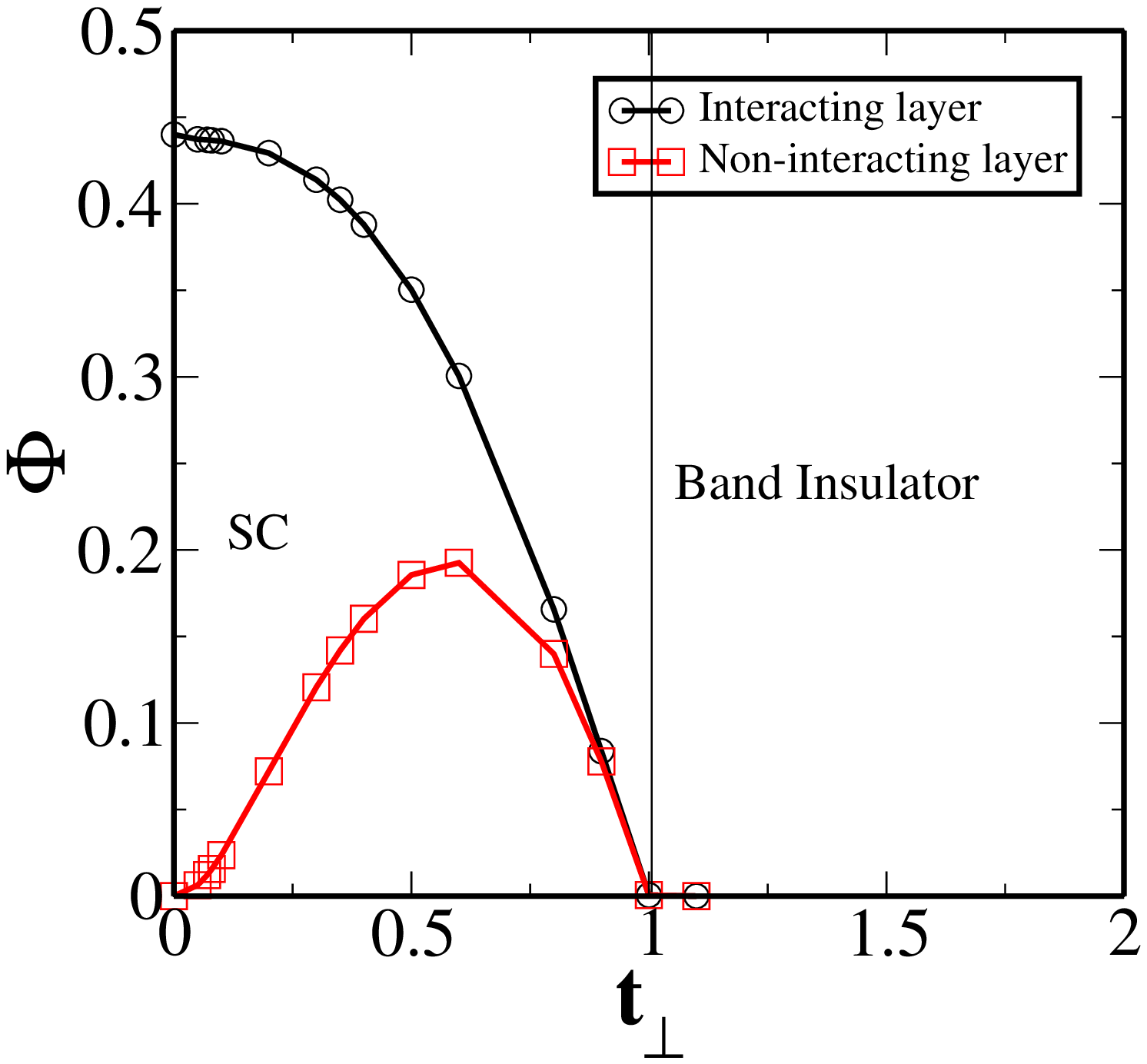}
\caption{ Order parameter -- Upper panel : IPTSC result; Lower panel :  BdGMF result, $\Phi$ vs $t_\perp$ at U=2.0, and at $\langle n_1\rangle=\langle n_2\rangle=1.0$.}
\label{fig:4fig3}
\end{figure}
\subsection{Superconducting order parameter $(\Phi)$ }
 Superconducting order is characterized by a finite value of $\Phi$, hence
this is a measure of  the strength of the pairing of electrons of opposite momentum and spin. The $\Phi$ for both planes is defined in equations~\ref{eq:4eq12} and~\ref{eq:4eq14}.
In the upper panel of figure~\ref{fig:4fig3}, $\Phi$ vs $t_\perp$ for both interacting  and non-interacting planes is shown. The 
$\Phi$ of interacting plane decreases monotonically with increasing $t_\perp$
 and beyond a critical value of $t_{c\perp}=t_\perp\sim0.6$, it completely vanishes and interacting plane becomes non-superconducting. In the 
 non-interacting plane, beyond $t_\perp\sim 0.08$, the $\Phi$ is non zero, indicating that superconductivity is induced in the non-interacting plane.
 Beyond $t_\perp\sim 0.08$, $\Phi$  increases with increasing $t_\perp$ and after a certain value of $t_\perp\sim 0.35$, $\Phi$ decreases with increases $t_\perp$ and beyond $t_{c\perp}=t_\perp\sim 0.6$, it completely vanishes, and non-interacting plane also becomes non-superconducting.

\subsection{Nature of the Spectral gap}
From figure~\ref{fig:4fig3}, since  $\Phi$ is finite for $t_\perp<
t_{c\perp}=0.6$, we infer that the nature of the spectral gap in figures~\ref{fig:4fig1} is superconducting. In the figures~\ref{fig:4fig2}, since $t_\perp >
t_{c\perp}$, the $\Phi$ is zero, and hence any gap would correspond to the
system being a simple
band insulator.

\subsection{Comparison of IPTSC and BdGMF Results }
To understand the effects of dynamical fluctuations over the static mean field, $\Phi$ vs $t_\perp$ within the BdGMF is shown in the lower panel of figure~\ref{fig:4fig3} for both interacting and non-interacting planes. Upper panel represents the  IPTSC result and the lower panel represents the BdGMF result. In both IPTSC and BdGMF framework, $\Phi$ for both interacting and non-interacting planes vanishes beyond a critical value of $t_\perp=t_{c\perp}$. This
critical value of $t_{\perp}$ hopping is higher in the BdGMF than in the IPTSC method. In the BdGMF framework, the superconducting phase continuously goes to insulating phase beyond  $t_{c\perp}$ and the intermediate metallic phase is not observed.
 While in the IPTSC framework, in both the interacting and non-interacting planes, the system first goes from superconductor to metallic phase and 
 subsequently with increasing $t_\perp$, both planes become insulating. Thus the inclusion of dynamical fluctuations strongly modifies the static mean field results.

\section{Conclusions}
In this letter, we have studied a bilayer attractive Hubbard model by combining DMFT and IPTSC at half filling. We have computed real frequency spectral functions and superconducting order parameter as a function of inter-layer hopping, $t_\perp$. Superconductivity is induced in the non-interacting layer due to proximity to superconducting plane, and beyond a critical value of $t_\perp$, both planes become non-superconducting. 
Our results are similar to a recent determinantal quantum Monte-Carlo (DQMC)
study~\cite{4Zujev} of a bilayer-AHM on finite-size lattices, in terms of the dependence 
of the order parameter on the $t_\perp$. However, it is not clear that,
the intervening metallic phase and the subsequent insulating phase shown
in figure~\ref{fig:4fig3},
are also found by the DQMC work.
 To understand the effect of dynamical fluctuations over static mean field, we have compared the superconducting order parameter computed using the IPTSC approach with that of the BdGMF approach. In the latter static
mean field approach, the intervening metallic phase is not observed, thus
implying that dynamical fluctuations play a very important role in the physics
of this SN bilayer system.
\acknowledgments
Authors thank CSIR, India and JNCASR, India for funding the research.\\\\
\author{\thanks{E-mail: \email{raja@jncasr.ac.in}}}

\end{document}